\documentclass[12pt]{article}
\usepackage{graphicx,amssymb,amsmath,amsthm}
\usepackage{cite}
\newcommand{\be}{\begin{equation}}
\newcommand{\ee}{\end{equation}}
\newcommand{\ben}{\begin{eqnarray}}
\newcommand{\een}{\end{eqnarray}}

\newcommand{\til}{\tilde}
\newcommand{\ii}{\'{\i}}
\newcommand{\ra}{\rangle}
\newcommand{\la}{\langle}
\newcommand{\kfo}{| x^o \ra}
\newcommand{\OA}{\hat{A}}
\newcommand{\OAP}{\hat{A'}}
\newcommand{\OAT}{\hat{A}^\dagger}
\newcommand{\kp}{|p \ra}
\newcommand{\kpt}{|\til{p}\ra}
\newcommand{\kpp}{|p' \ra}
\newcommand{\kn}{| n \ra}
\newcommand{\bn}{\la n |}

\newcommand{\lam}{\lambda}

\newcommand{\kfn}{| x_n \ra}

\newcommand{\oun}{\hat{U}}
\newcommand{\ouna}{\hat{U}^\dagger}
\newcommand{\kvq}{|q\ra}
\newcommand{\kvcp}{|p_c\ra}
\newcommand{\kvc}{|\til{p}\ra}
\newcommand{\kvatan}{|\eta_n\ra}
\newcommand{\bvatan}{ \la \eta_n|}

\def\R{\mathbb{R}}

\title{Statistical distribution,  
host for encrypted information} 
\author{L. Rebollo-Neira\\
Aston University\\Birmingham B4 7ET, United Kingdom
A. Plastino\\
Instituto de F\ii sica La Plata (IFLP)\\
Universidad Nacional de La Plata and CONICET
\thanks{Argentina's National Research Council}
\\C.C. 727, 1900 La Plata, Argentina}
\date{}
\begin{document}
\maketitle
\baselineskip = 1.7 \baselineskip

\begin{abstract}
The statistical distribution, when determined from an incomplete
set of constraints, is shown to be suitable as host for encrypted
information. We design an encoding/decoding scheme to
embed such a distribution with hidden information.
The encryption security is based on the extreme instability of the
encoding procedure. The essential feature of the proposed system
lies in the fact that the key for retrieving the code is generated
by random perturbations of {\em {very small value}}. The security
of the proposed encryption relies on the security to interchange
the secret key. Hence, it appears as a good complement to 
the quantum key distribution protocol.


\end{abstract}

\section{Introduction}
Cryptography is the art of code making and cryptology the art of
secure communications.  
Recently, quantum mechanics has made a remarkable entry in 
the field \cite{book1,book2,book3}.
The most straightforward application of
quantum cryptology is the distribution of secret keys. This
problem is refereed in the cryptography literature to as the key
distribution problem. Classical methods for securing a secret key
are based on the assumed difficulty of computing certain functions
\cite{bookc1,bookc2}. Quantum encryption provides a way of agreeing
on a secret key without making this assumption. 
The first quantum key distribution protocol was
proposed in 1984 by Bennett and Brassard \cite{bb84}
and there are already
rigorous proofs of its security \cite{BDVSW,shor,meyers}. 
The list of recent contributions concerning the  
security and implementation of quantum key distribution protocols 
is certainly extensive. 
Just as a sample one should mention \cite{ce1,ce2,ce3,ce4,ce5}.

The amount of information that can be transmitted
by a quantum transmission is not very large, but by means of
secret-key cryptographic algorithms a large amount of information
can be secured.
In this paper we set the foundations for a {\it statistical
distribution based encryption procedures}, which will be shown to
be good complements to quantum cryptology. It will be here
demonstrated  that the statistical distribution of a physical
system is a suitable host for encrypted information and we will
discuss a method for embedding encrypted messages without
affecting the physical content of the distribution.

The fact that a physical system can be assumed to be well
described by a particular parametric class of statistical
distribution entails, in most situations, the assumption of a
great deal of prior information. Indeed, the functional form of
most celebrated statistical distribution (Gibbs-Boltzmann,
Fermi-Dirac, Bose-Einstein distributions, etc) can be derived from
a few constraints expressed as mean values of some observable, and
the optimisation of a convex function called entropy or
information measure \cite{jaynes}. Thus, if a so obtained
distribution happens to be the right one to describe a particular
system, one can think of the optimisation process as replacing the
information required to determine a unique distribution for the
system.

We would like to think that, when one ``guesses'' (or derives) a
distribution from incomplete information, one also generates an
``invisible reservoir'' to place information. This is not an
original remark, of course, but an elemental result of linear
algebra: associated to a rank deficient transformation there are
two spaces, the {\em {range}} and {\em{null}} spaces of the
transformation. The latter is an ``invisible'' space in the sense
that all its elements are mapped to zero by the transformation.

In this Communication we show that the invisible reservoir is an
appropriate host for storing covert information. We present an
encoding/decoding scheme that allows to store a great amount of
hidden information as storing the distribution of a physical
system. The main idea is to make use of the null space of the
transformation generated by the constraints that should be
fulfilled by the distribution. The security of the system is
guaranteed by designing a (in the popular meaning of the vocable,
not in the technical one) ``chaotic'' encoding procedure. This is
achieved by means of random perturbations to an extremely
sensitive encoding process. The random perturbations of very small
values provide one thereby with the key for recovering the code.
This is the most remarkable feature of the our proposal: the key
for retrieving the hidden information is just a tiny number 
that accounts for the perturbation that has
been used for encoding purposes. Hence, the relevance of this
proposal in relation to quantum cryptology, and vice versa, since
the security of our proposal depends on a secure key interchange. 
It is appropriate to remark that the most 
notable difference between this approach and chaotic cryptosystems 
\cite{chao1,chao2}
is that the theory underlying our approach is essentially a linear 
one.

The idea of making use of an ``unstable system'' for
encryption has been successfully applied to over-sampling of
Fourier coefficients for transmitting hidden messages as
transmitting a signal \cite{mire}. Here we use equivalent ideas.
We assume that, in addition to some constraints, we have the
information on the process by which the statistical distribution
is univocally determined. This process is usually the optimisation
of a convex function (entropy). The particular expression for the
entropy may be a matter of controversy, though. For our purposes
the  choice of the entropic measure is not relevant at all.
What is important here is the convexity property 
to ensure a unique solution.
The selection of the appropriate entropic measure is
crucial, of course, to determine the right distribution for the
physical system. Nevertheless, this has no relation whatsoever
with our encryption scheme.

The paper is organised as follows: In section II we introduce the
notation together with
 an encoding/decoding scheme for
embedding a statistical distribution with hidden information. The
procedure is illustrated by a numerical simulation in section III
and some conclusions are drawn in section IV.
\section{Embedding the statistic distribution}
We restrict considerations 
to finite dimensional classical statistic systems, or 
equivalently, to a quantum system represented 
by a distribution constructed from commutative operators. 
In both cases the mean value of, say $M$, physical observations 
$x^o_1,x^o_2,\ldots, x^o_i,\ldots x^o_M$,
each of which is the expectation value
of a random variable that takes values
$x_{i,n} \;; \; n=1,\ldots, N$  according to 
a probability distribution
$p_n \; ;\; n=1,\ldots, N$ is expressed as: 
\ben
\label{con}
x^o_i&=&\sum_{n=1}^N p_n x_{i,n}
\;\;\;\;\; ;\;\;\;\;\;i=1,\ldots,M\nonumber\\
1&=& \sum_{n=1}^N p_n 
\een
Usually the number $M$ of available measurements 
is much less than the dimension $N$ of the probability space. 
In order to assert a unique distribution for the system at hand one 
has to adopt a decision criterion, which is frequently 
implemented through the maximisation of a convex measure 
on the probability distribution. Such a measure, 
called entropy or information measure, takes 
different forms.  
%Actually the adoption of a particular one 
%within some contexts  is 
%a subject of  controversy \cite{ , }.   The purpose of 
% this  Communication is far from that discussions though. 
Here we simple assume that the
distribution characterising a given physical system is 
agreed to be determined by a fixed set of constraints 
of the form \eqref{con} and the optimisation of a convex 
function that we denote $S$. 

For the sake of a handy notation we use Dirac notation 
to represent vectors. Thus,  
the probability distribution is represented as the ket 
$\kp \in {\R}^N$ which, 
by denoting as $\kn,\, n=1,\ldots,N$ the standard basis 
in ${\R}^N$, can be expressed as
\be
\kp = \sum_{n=1}^N \kn \bn p \ra= \sum_{n=1}^N p_n \kn.
\ee
We also define a vector $\kfo \in {\R}^M$ of components 
$x^o_1,x^o_2,\ldots, x^o_M,1$ 
and an operator $\OA:{\R}^N \to {\R}^{M+1}$ given by
\be
\OA = \sum_{n=1}^{N} \kfn \bn.
\ee
Vectors $\kfn \in {\R}^{M+1},\, n=1,\ldots,N$ are defined
in such a way that $\la i | x_n \ra =x_{i,n}, 
i=1,\ldots,M+1$  with $x_{M+1,n} =1$.  Hence, 
\be
\kfn= \sum_{i=1}^{M+1}|i \ra   \la i | x_n \ra
    = \sum_{i=1}^{M+1}  x_{i,n} |i \ra.
\ee
We are now in a position to joint the 
constraints \eqref{con} together in the equation 
\be
\label{consi}
\kfo= \OA \kp.
\ee
Since $\OA$ is a rank deficient operator, 
we know from elemental 
linear algebra that the general 
solution to the under-determined system \eqref{consi}
can be expressed in the form: 
\be
\label{geso}
\kp = {\OAP}^{-1} \kfo + \kpp,
\ee
where ${\OAP}^{-1}$ is the pseudo
inverse of $\OA$ (i.e. the inverse of the 
restriction of $\OA$ to  range($\OA$)) and $\kpp$ 
a vector in the null space of
that operator. Consequently, all the information the 
probability distribution contains concerning the 
data is expressible in the fashion
${\OAP}^{-1}\kfo$. On the contrary, 
the component $\kpp$ is completely independent
of the data, but strongly dependent on the 
selection criterion that is adopted to decide on one 
particular solution among the infinitely many solutions
the system \eqref{geso} has. 

The fact that all distributions of the form
$\kpt = {\OAP}^{-1} \kfo + \kpp$ with 
$\kpp \in$ null$(\OA)$ are capable of reproducing the 
constraints vector $\kfo$ provides us with a 
framework for the purpose of storing encrypted information
while storing the statistical distribution 
of a physical system.  At the encoding step we make 
the following assumptions:

\begin{itemize}
\item[ i)]The number $M+1$  of the independent linear equations which are used 
to determine the statistical distribution of a given system 
is fixed. The expected values  generating the equations 
are assumed to be known.  

\item[ii)]The probability distribution characterising the 
system arises by 
optimisation of a convex function $S$,  subjected to 
the $M+1$ linear constraints described above. 
\end{itemize}

Assumptions i) and ii) entail the availability, at this stage,  of vector $\kp$ and operator $\OA$. 
The vectors spanning the range and 
null spaces of this operator can be determined
by computing the 
the eigenvectors of operator 
$\hat{G}= \OAT \OA$. Let us  denote 
as $|\eta_n \ra,\, n=1,\ldots,N-(M+1)$
the normalised eigenvectors corresponding 
to zero eigenvalues.  We use these vectors to 
define operator 
$\oun : {\R}^N \mapsto {\R}^{N-(M+1)}$ as
\be
\oun=\sum_{n=1}^{N-M-1} \kn  \bvatan 
\label{oun}
\ee
This operator is termed {\em {decoding}} operator, and its 
adjoint, $\ouna$,
{\em {encoding}} operator.
Using  $\ouna$
a basic code of
$N-(M+1)$ numbers is constructed as follows:
Let the $N-(M+1)$numbers be the $N-(M+1)$-components
$\la n \kvq =q_n\,, n=1,\ldots,N-(M+1)$ 
of vector $\kvq\in {\R}^{N-(M+1)}$
and define:
\be
\kvcp=  \ouna \kvq= \sum_{n=1}^{N-M-1} \kvatan \la n \kvq.
\label{hi1}
\ee
Given a distribution $\kp$, amenable to be 
determined from the optimisation of an entropy measure $S$ and 
a set of $M+1$ constraints, the code $\kvq$
is embedded in the distribution through the process below.

\subsection*{Encoding process}
\begin{itemize}
\item Compute vector $\kvcp$ as in (\ref{hi1}).
\item Add $\kp$ and $\kvcp$ to construct
$$\kvc = \kp + \kvcp.$$
\end{itemize}
\subsection*{Decoding process}
\begin{itemize}
\item
Use the vector $\kvc$ to recover the data $\kfo$ as
$$\kfo = \OA \kvc$$
\item
Use the data $\kfo$ to determine, by optimisation of 
 $S$, the distribution $\kp$. 
From $\kvc$ and $\kp$ compute vector
$$\kvcp = \kvc -\kp.$$
\item
Use the decoding operator $\oun$ to obtain the encrypted code 
by noticing that, since $\oun \ouna$ is the identity 
operator in ${\R}^{N-(M+1)}$, from (\ref{hi1}) one has:
\be
\kvq=  \oun \kvcp.
\ee
\end{itemize}
Note that the success of the above encoding/decoding scheme 
relies on the possibility of ordering the eigenvectors
in the null space. This is perfectly possible by fixing 
the numerical method for computing 
the eigenvectors of $\hat{G}$.
However, the process is extremely 
unstable, as a {\em{tiny perturbation}}
to any of the matrix elements of 
operator $\hat{G}$ produces a {\em{huge effect}}
in the eigenvectors of zero eigenvalues. 
This ``chaotic'' (in the popular sense) behaviour
of the eigenvectors in the null space provides,
 naturally, the security key for retrieving
the encrypted code. Indeed, consider that $\epsilon$ 
is a very small number (order of $10^{-13}$, say) that 
we add to one of the matrix elements 
of $\hat{G}$. Such a tiny perturbation 
 does not yield any 
detectable effect in the reconstruction of the distribution 
$\kp$ but an enormous effect with regard to the eigenvectors  
in the null space.
Hence, as illustrated by the examples of the next section, 
{\em the perturbation $\epsilon$ provides the safety 
key of our encoding/decoding scheme}. 

\subsection{Numerical Examples} 
Consider that a probability distribution concerning 
an event space of dimension $N=401$ is appropriately 
determined, by the Jaynes maximum entropy 
formalism,  
from the normalisation to unity constraint and 
the first four  moments of a 
random variable $x_n, n=1,\ldots,401$ that takes 
 values ranging from $x_1=-1$ to $x_{401}=1$ 
 with uniform increment 
$\Delta=1/400$. Thus 
the distribution, which arises by maximisation of the 
Shannon's entropy 
$S= -\sum_{n=1}^{401} p_n \ln{p_n}$ 
subjected to the given constraints, has the form:
\ben
p_n&=&e^{-\lam_o - \lam_1 x_n - \lam_2 x_n^2 
- \lam_3 x_n^3 -\lam_4 x_n^4}\nonumber\\
e^{\lam_o}&=& \sum_{n=1}^{401} e^{-\lam_1 x_n - \lam_2 x_n^2 
- \lam_3 x_n -\lam_4 x_n^4}.
\een
The parameters $\lam_1, \lam_2, \lam_3$  and 
$\lam_4$ are determined from the equations
\be
x_i^o= \frac{\sum_{n=1}^{401} x_n^i e^{-\lam_1 x_n - \lam_2 x_n^2
 - \lam_3 x_n^3 -\lam_4 x_n^4} }{
\sum_{n=1}^{401} e^{-\lam_1 x_n - \lam_2 x_n^2
-\lam_3 x_n^3 -\lam_4 x_n^4}}, \;\;\;\; i=1,\ldots,4
\ee
For $x_1^o= -0.0224, x_2^o=  0.1048 , x_3^o = 
 -0.0124, x_4^o =  0.0284 $ one obtains 
$\lam_1=-0.3, \lam_2=3, \lam_3=2, \lam_4=3.8$.
The operator $\OA$ has a $5\times 401$ 
matrix representation of elements $x_{i,n}=x_n^i, 
i=1,\ldots,4;\,n= 1,\ldots,401$ and 
$x_{5,n}=1, n=1,\ldots, 401$. We construct 
operator $\hat{G}= \OAT \OA$ and compute its 
eigenvectors.  The 396 eigenvectors
corresponding to zero eigenvalues are used to construct   
the encoding operator $\ouna$  
to encrypt a code $|q\ra$ of $396$ numbers. 
These numbers, each of which consists of 15 digits, are 
taken randomly from the $[0\,,\,1]$ interval.
We now proceed as indicated in the encoding 
process of the previous section:  We construct 
the vector $ \kvcp= \ouna |q \ra$   and 
add it to the  distribution 
$\kp$ to obtain 
vector $\kvc = \kp + \kvcp$. This vector contains 
 both,  the information  on  the 
physical system and the code. In order to 
retrieve  such information we use $\kvc$ to generate 
the constraints as $x_i^o = \la i |\OA \kvc,\, i=1,\ldots,5$. 
Since, by construction, $\OA \kvcp =0$, the constraints are 
generated from $\kvc$ with high precision. We use then this values 
to solve for the parameters of the distribution so us to recover 
 $\kp$. Vector $\kp$  allows to 
obtain vector $\kvcp$ from the available vector 
$\kvc$ as  $\kvcp=  \kvc -\kp$. 
The code is  thus retrieved 
by the operation $|q\ra=\ouna \kvcp$. 
Table 1 gives  five  of the 396 
code numbers. The second  column 
corresponds to the reconstructed numbers.  
As can be observed the quality of the 
reconstruction is excellent. 
In order to give a measure assessing
the reconstruction of all numbers,
let us denote  by $|q^{\rm{r}}\ra$
the reconstructed code  and  define
the error of the reconstruction as
$\delta^{\rm{r}}=|| {\kvq} -|q^{\rm{r}}\ra||.$
The value of $\delta^{\rm{r}}$ is in this case 
$4.8 \times 10^{-14}$.

Let us now  distort the matrix 
representation of operator $\hat{G}$ by adding 
a number $\epsilon= 2.9\times 10^{-13}$ to one of its  
elements, say the element at the 
first row and fifth column.  
If we repeat the process using the
distorted matrix the outcomes are the 
following: The perturbation has no detectable 
effect in the reconstruction of the 
distribution $\kp$. However,  
if we intend to reconstruct the code without considering
the perturbation, what we obtain has no
relation whatsoever with the true code 
(see the 3rd column of Table 1). 
The error of the reconstruction is 
$\delta^{\rm{r}}=17.25$.  
Since the recovery of the code is
{\em{only}} possible
if the value of the perturbation is known, 
the key for recovering the code is the
value of the perturbation and the to numbers 
 labelling the  
element that has been distorted, in this case 
$(1,5)$. Of course, rather than 
distorting one matrix element we may wish to distort a 
random number of them. In such a case the 
decoding key becomes a string of ordered 
pairs of natural numbers, indicating the elements that were 
randomly selected to be distorted and 
the corresponding values of the perturbations. 
Moreover, to avoid attacks of the type
{\em known plaintext attack} \cite{bookc1,bookc2},
in which the attacker is supposed to have
collected correctly decrypted message in order to use them to
decrypt others, one can proceed as follows: maintaining
one perturbation secret as the key for decryption,  {\it other}
perturbations are made public and are different for
every message. This  avoids the repetition of
the encoding operator with the same key. Thus, the
knowledge of decrypted messages does not provide
information on the encoding operator to
encrypt other messages with the same key. This prevents
thereby the  possibility of known plaintext attacks.

We would like to stress that the success of the 
proposed procedure strongly depends on the use of  
the identical numerical method to obtain 
the same basis of null($\hat{G}$) in the encoding and 
decoding process. Nevertheless, the procedure does 
not depend on the machine processor. 
In the examples presented here the encoding 
was performed in powerful computer cluster, 
and the decoding in a laptop, 
using Matlab 6.5. 
Let us also remark that only the precision  
in the representation of
operator $\hat{G}$ is crucial for the code reconstruction, 
since the numerical errors in determining $\kvcp$  
are not magnified. This is 
due to the fact that, since  $\ouna \oun$ is the 
identity operator in ${\R}^{N-(M+1)}$, the inverse 
recovering of $|q \ra $ from  $\kvq=  \oun \kvcp$ is
very stable against perturbations of $\kvcp$.\\

\begin{table}
\begin{center}
\subsection*{}
\begin{tabular}{||c|  c| c ||}  \hline
Code numbers &  reconstruction  & disregarding perturbation\\
\hline
 0.43596704982551 & 0.43596704982551& -0.04525907549619\\
 0.82124392828478 & 0.82124392828478 & 0.18665860768833\\
 0.66471591347310 & 0.66471591347310& 0.19297513576254\\
 0.64554449242809 & 0.64554449242809 & 0.08172338626606\\
 0.84398334243978 & 0.84398334243978 & -0.14994519073447\\
\hline
\end{tabular}
\end{center}
\caption{5 code numbers and their
reconstruction. Assuming the perturbation  to 
be known (second column) and otherwise (third column)}
\end{table}

\section{Conclusions}
An encoding/decoding scheme for embedding hidden information into
the statistic distribution of a physical system has been
presented. The encryption security is based on the extreme
instability of the encoding process, which is endowed with the
following feature: a tiny perturbation to the matrix yielding the
eigenvalues used to construct the encoding operator produces a
huge effect in the recovery process.

Thus, the key for retrieving the code is given by the value of
the perturbation. The security for interchanging the key is, of
course, essential, but we can rely on the secure quantum key
distribution protocol to ensure a safe key delivery. Conversely,
the quantum protocol can make use of the proposed setting for
encryption, as it entails the transition of very little
information through the quantum channel.

A remarkable property of making use of a statistical distribution
for the purpose of storing encrypted information  is the fact
that, the larger the dimension of the distribution is, the larger
the amount of  encrypted information that can be stored.  This
opens the possibility of devising
more sophisticated encryption algorithms than the one advanced here,
yet based on the same principles.

We believe that the results we have presented are certainly
encouraging and feel confident that they will  contribute to many
fruitful discussions and follow-up work in the subject. Finally we
would like to stress that the proposed scheme is not restricted to
be applied only on physical
distributions.
When a physical distribution is involved, one has also stored 
the information on the physical system. Hence the importance of 
using an appropriate entropic measure 
for the encoding/decoding process. If the
entropic measure is the right one, after recovering the
statistical distribution it can be used to make correct
predictions on the expected values of physical quantities which
are not experimentally available.

%\begin{references}

\end{document}